\documentstyle[12pt]{article}
\input epsfig.sty
\def\beq{\begin{equation}}
\def\eeq{\end{equation}}
\def\bea{\begin{eqnarray}}
\def\eea{\end{eqnarray}}

\def\NP{{\it Nucl. Phys.} }
\def\PL{{\it Phys. Lett.} }

\def\ap{\alpha ^{\prime}}
\begin {document}
\begin{titlepage}
December 1997 \\
\begin{flushright}
HU Berlin-EP-97/89\\
\end{flushright}
\mbox{ }  \hfill hepth@xxx/9712057
\vspace{5ex}
\Large
\begin {center}
\bf{Non-Abelian gauge field dynamics on matrix D-branes
in curved space and two-dimensional $\sigma$-models${}^\dagger$}
\end {center}
\large
\vspace{1ex}
\begin{center}
H. Dorn
\footnote{e-mail: dorn@qft3.physik.hu-berlin.de}
\end{center}
\normalsize
\it
\vspace{1ex}
\begin{center}
Humboldt--Universit\"at zu Berlin \\
Institut f\"ur Physik, Theorie der Elementarteilchen \\
Invalidenstra\ss e 110, D-10115 Berlin
\end{center}
\vspace{4ex}
\rm
\begin{center}
{\bf Abstract}
\end{center}
A survey is given of the formulation of a $\sigma$-model describing an open
string moving in general target space background fields and coupling to both 
a matrix-valued D-brane position and a 
matrix-valued gauge field on the D-brane. The equations of motion for
the D-brane and the gauge field are derived from the conformal invariance 
condition on the string world sheet in lowest order of $\ap$. The ordering
problem of the involved matrices is solved.\\ 
In addition to our previous work we 
discuss a conflict between the classical T-duality rules and renormalization.
The calculation of the RG $\beta$-functions does not yield the mass term
obtained by formal application of these rules in the case of target space separated D-brane copies.
\vfill      \hrule width 5.cm
\vskip 2.mm
{\footnotesize
\noindent $^\dagger$Talk given at the workshop 
``Quantum Aspects of Gauge Theories, Supersymmetry and Unification'',
Neuch\^{a}tel, September 18-23, 1997, to appear in the proceedings.}   
\end{titlepage}
\setcounter{page}{1}
\pagestyle{plain}
\section{Introduction}
Dirichlet branes appear via T-duality transformations of open strings and,
independently, they are necessary in type II theories to complete the duality
pattern between the various string theories, see e.g.\cite{polrev}. In this
contribution we will discuss the situation of an open string propagating
in nontrivial target space fields
$G_{\mu\nu}(X),~B_{\mu\nu}(X),~\Phi (X)$.\footnote{$B\neq 0 $ in type II 
theories only.}
Its endpoints are bounded to a D-brane whose target space position is described
by $X^{\mu}=f^{\mu}(Y)$, where $Y^M,~ M=1,...,p+1,$ are coordinates on the 
D-brane. The end points can move freely within the D-brane and couple to
an Abelian gauge field $A_M(Y)$ living on the D-brane. Here we do not take into 
account any RR-background field. The equations of motion for $G,~B,~\Phi ,~A$
and $f$ can be derived from the conformal invariance condition of the 
two-dimensional $\sigma $-model living on the world sheet of the string.
The resulting equations of motion are equivalent to the stationarity 
condition of the Born-Infeld action reduced to the D-brane \cite{leigh}
\bea
S_{BI}~=~\int ~d^{p+1}Y~\mbox{e} ^{-\phi (Y)}\sqrt{\mbox{det}(g+b+2\pi\ap F)}~.
\label{1}
\eea
Here $g,~b,~\phi$ are the fields induced on the brane by $G,~B,~\Phi $. $F$
is the field strength to $A$.

The generalization to non-Abelian gauge fields on the D-brane is a subtle
issue within the $\sigma$-model approach. Part of the problem is present
even before taking into account D-branes. A non-Abelian structure is 
introduced by 
attaching Chan-Paton indices to the endpoints of the string. Now in general
the partition function $Z[G,B,\Phi ,A]$ should be an off shell extension
of the generating functional of string scattering amplitudes. 
Due to the index structure connected with the emission of an open string 
there is no such $\sigma$-model for
a particular $(a,b)$ excitation for $a\neq b$.
Indeed if an $(a,b)$ string emits an $(a,c)$ string an $(c,b)$ string 
remains. There is no choice of $c$ possible such that all three participants
are of the same type. The summation over the allowed values
of the Chan-Paton indices on all boundary components of the world sheet for
N-point amplitudes and summation over N yields the Wilson loop for the gauge 
field introduced as a source of the gauge excitations of the 
string.\footnote{For a somewhat different approach see \cite{mav}.} 
One should stress that the boundary of the Wilson loop $\sigma $-model carries a Chan-Paton double index. This is in contrast to the world sheet boundary
of a particular $(a,b)$ string, as can be seen in the figure. 
\begin{figure}[t]
\begin{center}
\mbox{\epsfig{file=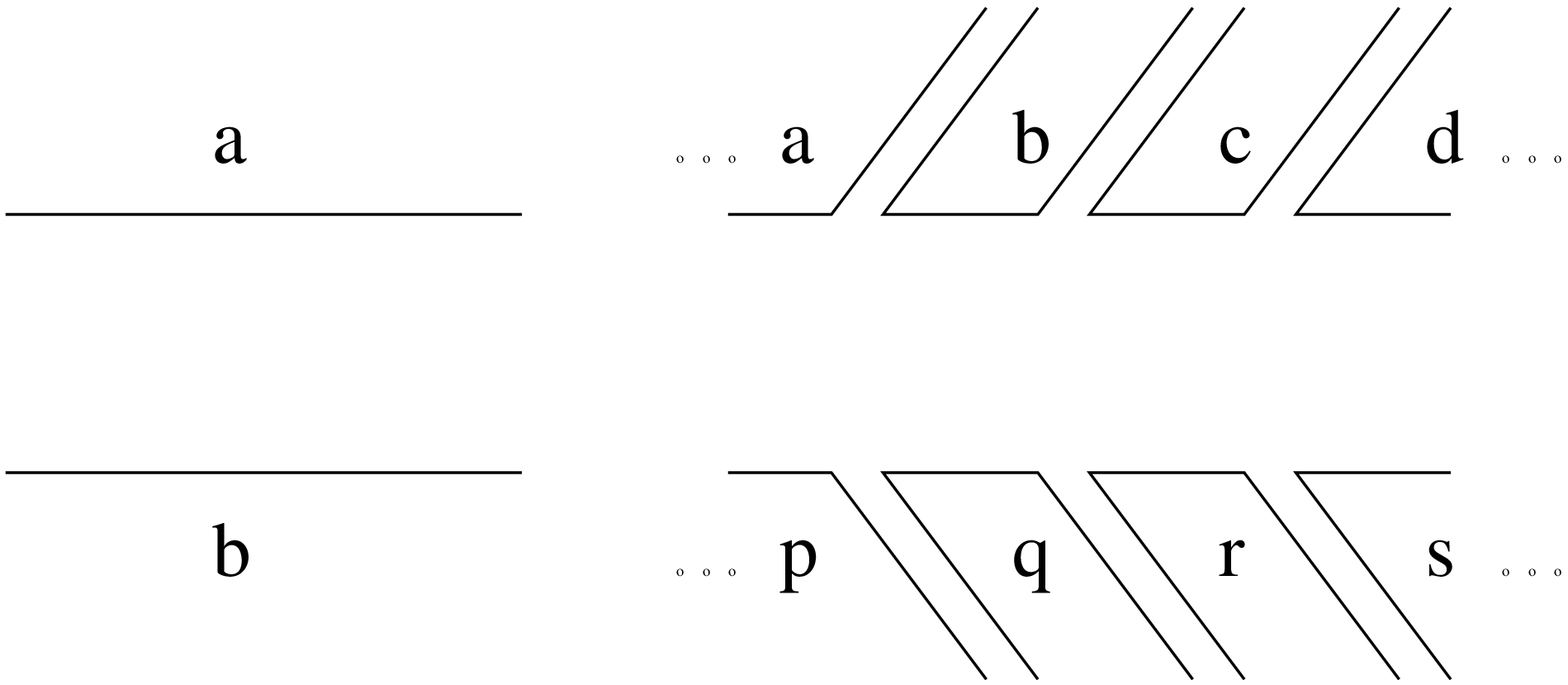, width=70mm}}
\end{center}
\noindent {\bf Fig.}\ \ {\it
Chan-Paton index structure of the boundary of the 2-dim. manifold for\\
$~~~~~~~~$an (a,b) string and for the Wilson loop $\sigma $-model.}
\end{figure}
The Wilson loop along the world sheet boundary $\partial M$ is given by 
$\mbox{tr}U$ with 
$U[\partial M,A]=P\exp (i\int A_{\mu}dX^{\mu})$ and has, due to the path
ordering $P$, not
the structure of the exponential of a local boundary action as in the Abelian 
case. Therefore, the analog of the Lorentz force term in the boundary
condition of a string with freely moving ends appears to be
$\propto \mbox{tr} P(F_{\mu\nu}\dot X^{\nu}~U[\partial M,A])$, which is a
nonlocal term. There is however a well known trick to achieve a local 
description at least in intermediate steps of the calculation. Introducing
a 1-dimensional auxiliary field living on $\partial M$ with propagator
$\langle \bar{\zeta}_a (s)\zeta _b (s^{\prime})\rangle _0=\Theta (s^{\prime}
-s)\delta _{ab}$ one can write 
\cite{zform}
\beq
U_{ab}[\partial M,A]~=~ \int _{\partial M}D\bar{\zeta}D\zeta ~\bar{\zeta}_b(0)
\zeta _a(1) \exp \left ( -\int _0^1(\bar{\zeta}\frac{d}{ds}\zeta (s)-
i\bar{\zeta}A_{\mu}(X(s))\zeta \dot X^{\mu})~ds\right )~.
\label{2}
\eeq
Under the $\bar{\zeta},\zeta$ auxiliary integral the Lorentz force now
is $\propto \bar{\zeta}(s)F_{\mu\nu}(X(s))\zeta (s) \dot X ^{\nu}(s)$.

Let us take aside for a moment the aim to find a $\sigma$-model picture.
Then the generalization to the non-Abelian situation of the model for an open 
string bound to some D-brane can proceed along the following line of arguments 
\cite{witten,polrev}. One introduces as many D-branes as the Chan-Paton index 
takes values ($a=1,2,...N$). The string endpoint with index $a$ has to sit on 
the D-brane copy number $a$. As long as the various copies are lying on 
top of each other the situation is, concerning the gauge field excitations, 
simply the same as for a string with free ends in a reduced space. The
gauge fields live on the D-brane. As a second step one allows the D-brane 
copies to be separated in target space. Then due to the string tension 
gauge field excitations $(a,b)$
become massive if the branes with number $a$ and number $b$ are separated. 
The $N$ brane positions are interpreted to be the entries in a 
diagonal $N\times N$ matrix. Finally the D-brane position is allowed
to become a full $N\times N$ matrix. This last step which puts the
now matrix-valued brane position and the gauge field living on the
brane into the same gauge structure is motivated mainly, at least to my 
understanding, by a look at the formal extension of the T-duality
rules for open strings in Abelian gauge field background. In that
case the components of the gauge field in the isometry directions become
the D-brane position in the dualized version. There is also a possibility
to relate the nondiagonal matrix elements to excitations of strings
spanned between different D-brane copies directly \cite{doug}.

We summarize in section 2 the outcome of our functional integral treatment
of T-duality of the $\sigma$-model describing an open string coupled to
non-Abelian gauge fields based on the use of the auxiliary $\zeta$-field 
formalism \cite{do,d}. It gives a meaning to the notion of matrix valued
D-brane position suitable for practical calculations. The model obtained after
T-dualizing allows a straightforward generalization to the generic case
without any isometry. Thereby the ordering problem for the target space
fields at matrix-valued arguments among themselves as well as with the
gauge field and the D-brane position is solved. To demonstrate the
calculational power of the formalism the result of the lowest order in $\ap $
calculation of the gauge field and brane position RG $\beta$-function is 
presented.

The final section 3 is devoted to the problem of mass term generation
for the situation of separated D-brane copies discussed above on the 
heuristic level. We show how a formal application of the T-duality
rules to the standard Yang-Mills equation of motion yields such a term.
On the other side we find no hint to such a term in the direct 
$\beta$-function calculation in the D-brane model. 
\section{$\sigma$-model construction and its RG $\beta$-functions}
The partition function for the $\sigma$-model describing an open string with
free ends coupling to a non-Abelian gauge field $A_{\mu}(X)$
is given by ($\Psi $ is a collective notation for $G,~B,~\Phi $. $\gamma $ is
a metric on the string world sheet.)
\beq
Z[\Psi ,A]~=~\int DXD\bar{\zeta}D\zeta ~\exp\left (-\int_0^1\bar{\zeta}\dot
{\zeta}~ds~+~iS[\Psi ,C;X]\right )~,
\label{3}
\eeq
with
\bea
S[\Psi ,C;X]&=&S_M[\Psi ;X]~+~\int_{\partial M} \left (C_{\mu}(X(z(s)),s)\dot 
X^{\mu}-\frac{1}{2\pi}
k(s)\Phi \right )ds~,\nonumber \\
S_M[\Psi ;X]&=&\frac{1}{4\pi \alpha ^{\prime}}\int _M d^2 z \sqrt{-\gamma}
\left ( \partial _m X^{\mu} \partial _n X^{\nu} E^{mn}_{\mu \nu}(X(z))+
\alpha ^{\prime} R^{(2)}\Phi (X(z)) \right )~,\nonumber\\
E^{mn}_{\mu \nu}(X)&=&\gamma ^{mn}G_{\mu \nu}(X)~+~\frac{\epsilon ^{mn}}{\sqrt
{-\gamma}}
B_{\mu \nu}(X)~,\nonumber\\
C_{\mu}(X(z(s)),s)&=&\bar \zeta (s)A_{\mu}(X(z(s)))\zeta (s)~.
\label{4}
\eea
We now interchange the order of integrations over the $\zeta$-auxiliary field 
and the string position $X$. Then before performing the final
$\zeta$-integration we find ourselves in an Abelian situation. 
The only difference to the
genuine Abelian case is an explicit boundary parameter $(s)$ dependence of 
the gauge field $C_{\mu}$.
The standard manipulations to derive Buscher's rules \cite{bu} within the 
functional integral formalism can now be repeated. For the simplest case
of one isometry $X^{\mu}=(X^1,X^M),~\partial _1\Psi =\partial _1A=0$ one 
gets \cite{do,d}
\beq
Z~=~ \int D\bar{\zeta}~ D\zeta ~\bar{\zeta}_b (0)\zeta _b (1)\mbox{e}^{
-\int_0^1\bar{\zeta}\dot{\zeta}~ds}~{\cal F}[\tilde{\Psi},\bar{\zeta}
\tilde A \zeta \vert -2\pi\ap\bar{\zeta}A_1
\zeta ]~.
\label{5}
\eeq
$\cal F$ is a functional of the target space fields and a function $f$
specifying the boundary condition. It is defined by
\beq
{\cal F}[\Psi,C\vert f]~=~\int _{X^1(z(s))=f(X^M(z(s)),s)} D X~\exp (iS
[\Psi ,C;X])~.
\label{6}
\eeq
The dual target space fields are given by the standard (closed string)
Buscher rules for $\tilde{\Psi}\leftrightarrow \Psi $ as well as by
\beq
\tilde A_{\mu}=(0,A_M)~,~~~~~\mbox{with}~~\tilde{X}^M=X^M~.
\label{7}
\eeq
The boundary condition in (\ref{6}) is an explicit $s$-dependent Dirichlet
condition. Matrices appear as arguments in $\cal F$ only sandwiched between
$\bar{\zeta}$ and $\zeta $. The final $\zeta $-integration undoes sandwiching
and orders the matrices with respect to their corresponding boundary parameter
value. This ordering then concerns matrices appearing in the gauge field
argument of $\cal F$ as well in its boundary condition specifying argument.
Thereby the notion of matrix valued boundary condition gets a meaning suitable
for practical calculations.

Before the final $\zeta $-integration there is a simple D-brane interpretation
of the boundary condition as follows. In the Abelian gauge field case
there is an a priory\footnote{Neglecting for a moment the D-brane dynamics.}
given manifold $\tilde{X}^1=f(\tilde{X}^M)=-2\pi \ap A_1(X^M)$. The boundary 
of the 
string world sheet has to be situated on this manifold. In the non-Abelian case
there is an infinite set of manifolds depending on one real parameter
$\tilde{X}^1=f(\tilde{X}^M,s)=-2\pi\ap \bar{\zeta}(s)A_1(X^M)\zeta (s)$. The 
string
world sheet boundary point $\tilde{X}^{\mu}(z(s))$ has to sit on the manifold
corresponding to the parameter values $s$, i.e. $\tilde{X}^{\mu}(z(s))=
(f(\tilde{X}^M(z(s)),s),\tilde{X}^M(z(s)))$.
Perhaps it is useful to stress that the manifolds for different value of $s$
should not be confused with the finite number of D-brane copies discussed in 
the introduction.

As long as one considers the D-brane model in the presence of isometries  
just derived via T-duality there is no need to handle the target space
fields $\Psi $ at matrix valued arguments because the fields do not depend
on the corresponding coordinates. However, to define a non-Abelian D-brane
model in the generic non-isometric situation one has to give a prescription
for handling $\Psi $ at matrix valued D-brane positions. Then a straightforward
definition of the $\sigma $-model is achieved by allowing dependence
on all coordinates in eqs.(\ref{5},\ref{6}). For a $(p+1)$-dimensional D-brane
with matrix valued position $f^{\mu}(Y)$
\footnote{We drop the tilde and do not introduce a new letter to distinguish 
the matrix 
$f^{\mu}(Y)$ and its sandwiched version $f^{\mu}(Y(s),s)=\bar{\zeta}(s)
f^{\mu}(Y(s))\zeta (s)$.} and a non-Abelian gauge field $A_M (Y)$ this leads
to the partition function 
\beq
{\cal Z}[\Psi ,A]~=~ \int D\bar{\zeta}(s) D\zeta (s) ~\bar{\zeta}_b (0)\zeta _
b (1)e^{-\int_0^1\bar{\zeta}\dot{\zeta}~ds}~{\cal F}[\Psi,\bar{\zeta}A \zeta
\vert \bar{\zeta}f\zeta ]~,
\label{8}
\eeq
with 
\beq
{\cal F}[\Psi,\bar{\zeta}A\zeta \vert \bar{\zeta}f\zeta ]~=~
\int _{X^{\mu}(z(s))=\bar{\zeta}(s)
f^{\mu}(Y(s))\zeta (s)}DX\exp (i{\cal S}[\Psi,A;\bar{\zeta},\zeta ;X])~,
\label{9}
\eeq
\bea
{\cal S}[\Psi,A;\bar{\zeta},\zeta ;X]&=&S_M[\Psi ;X]~+~{\cal S}_{\partial M}~,
\nonumber\\
{\cal S}_{\partial M}&=&\int _{\partial M}\left
(\bar{\zeta}(s)A_N(Y(s))\zeta (s)\dot{Y}^N-
\frac{1}{2\pi}k(s)\Phi (X(z(s)))\right )ds~.
\label{10}
\eea    
The RG $\beta $-functions related to the D-brane fields have been calculated
in \cite{d} in lowest order of $\ap $.
Defining an operation ${\cal Q}$ by
\beq
{\cal Q}\{\prod _{j=1}^N\left (\bar{\zeta}{\cal M}_j\zeta \right )\}
~=~\mbox{Sym}\left (\prod _{j=1}^N {\cal M}_j \right )~.
\label{11}
\eeq 
the result can be summarized as (for $\Phi =0$)
\bea
\beta ^{(A)}_L&=&{\cal Q}\{ g^{MN}(\hat D_Mb_{NL}+2\pi\alpha ^{\prime}\bar
{\zeta}\hat D_MF_{NL}
\zeta -i\bar{\zeta}[F_{LM},f^{\alpha}]\zeta ~f^{\nu}_{;N}~B_{\nu\alpha})
\}~,\nonumber\\
\beta ^{(f)}_{\nu}&=&{\cal Q}\{ g^{MN}K_{MN}^{\mu}G_{\mu\nu}+\frac{1}{2}(b^{MN}+2
\pi\alpha
^{\prime}\bar{\zeta}F^{MN}\zeta )f^{\alpha}_{;M}f^{\beta}_{;N}H_{\alpha\beta
\nu} \} ~.\label{12}
\eea
$H$ is the field strength related to $B$,
\\$g_{MN}(Y(s),s)=f^{\mu}_{;M}(Y(s),s)\cdot f^{\nu}_{;N}\cdot G_{\mu \nu}
(\bar{\zeta}(s)f(Y(s))\zeta (s))$, $D_M$
is the gauge covariant derivative, $\hat D_M$ the total covariant derivative
and $K^{\mu}_{MN}$ the covariant external curvature.
Note that ${\cal Q}$ acts on all the sandwiched matrices either  
written down in eq.(\ref{12}) explicitly or appearing as arguments of 
$g,\hat{D},b,B,H,K,f^{\nu}_{;N}$.

In connection with the discussion of the next chapter one should add, 
that in \cite{d} there is a simplified treatment of the variational formula
for the Wilson loop in $\zeta $-language. Meanwhile I have performed the 
complete procedure taking into account the appearance of the $\zeta $-field 
in the boundary condition, too. However, the final result (\ref{12}) remains
unchanged.   
\section{The mass term}
The standard functional integral manipulations to derive the T-duality
rules are of formal nature since they do not take care of renormalization
effects and functional determinants. Therefore, one has to check whether the 
derived rules indeed relate equivalent two-dimensional $\sigma $-models 
as quantized field theories on the string world sheet. For $\sigma $-models
without boundary the Buscher rules get corrections at 2-loop order 
\cite{haage}. In the presence of boundaries coupling to Abelian gauge
fields the Buscher rules supplemented by the Dirichlet condition 
\footnote{In this section we use the tilde to denote the fields in the
free end model.}
\beq
X^1~=~f(X^M)~=~-2\pi\ap \tilde A_1(\tilde X^M)
\label{13}
\eeq
are respected for constant field strength in all orders of $\ap $ (For a 
discussion see e.g. second ref. in \cite{do}.).
Let us look at this question for non-Abelian gauge field in the simplest case
of Minkowskian target space with $B=0~,\Phi =0$. Then (\ref{12}) reduces to
\beq
\beta ^{(A)}_L~=~2\pi\ap {\cal Q}\left (g^{MN}\bar{\zeta}D_MF_{NL}\zeta 
\right )~,~~~~~~~~~~\beta ^{(f)}_{\nu}~=~\eta _{\mu\nu}{\cal Q}\left (
g^{MN}K^{\mu}_{MN}\right )~,
\label{14}
\eeq
with
$$g_{MN}~=~\bar{\zeta}D_Mf^{\mu}\zeta ~\bar{\zeta}D_Nf^{\nu}\zeta ~\eta _
{\mu\nu}~.$$

On the other side, applying the rule (\ref{13}) as well as $\tilde X^M=X^M,~
A_1=0,~ A_M=\tilde A_M$ to the Yang-Mills $\beta $-function  for an open 
string with free ends
$$\tilde{\beta}^{(\tilde A)}_{\lambda}~=~2\pi\ap ~\eta ^{\mu\nu}~\tilde{D}
_{\mu}\tilde{F}_{\nu\lambda}$$
we get
\bea
\tilde{\beta}^{(\tilde A)}_L&=&2\pi\ap ~\eta ^{MN}~D_MF_{NL}~-~\frac{i}
{2\pi\ap}\left [f,D_Lf\right ]~,\nonumber\\
\tilde{\beta}^{(\tilde A)}_1&=&-\eta ^{MN}~D_MD_Nf~.
\label{15}
\eea
Besides the subtlety that the induced metric in (\ref{14})
$g_{MN}=\eta _{MN}+\bar{\zeta}D_Mf\zeta ~\bar{\zeta}D_Nf\zeta ~\eta _{11}$
differs from $\eta _{MN}$ the most striking difference between
(\ref{14}) and (\ref{15}) is the appearance of the second term for
$\tilde{\beta}^{(\tilde A)}_L$ in (\ref{15}).
In the spirit of the heuristic discussion of the first section such a
term is highly welcome. It can be interpreted as a mass term for the 
non-Abelian gauge field. This becomes obvious at least in the special D-brane
configuration of diagonal and constant $f$, i.e. $f_{ab}=f_a\delta _{ab},~
\partial _Mf_a=0$, which describes planar parallel D-brane copies at position
$X^1=f_a,~a=1,...,N$. The equation of motion obtained from 
$\tilde{\beta}^{(A)}_L=0$ then is  
\beq
(D^M~F_{ML})_{ab}+\left (\frac{1}{2\pi\ap}\right )^2 (f_a-f_b)^2 (A_L) _{ab}~=~0~.
\label{16}
\eeq
As expected the mass is proportional to the separation of the D-brane copies.
Perhaps it is useful to point out that the mass term in its general form
in (\ref{15}) is gauge invariant.

Due to its $(\ap)^{-1}$ power the mass term in (\ref{15}) by no means can arise
in the direct $\beta $-function calculation of the $\sigma $-model defined
by (\ref{8})-(\ref{10}). Such a calculation for sure yields non-negative
powers of $\ap $ only. Therefore we have a problem independent of the details
of the calculation in \cite{d}. So far we considered the RG $\beta $-function
which may differ from the Weyl anomaly coefficients. But the 
argument based on $\ap $ power counting applies to possible radiative 
corrections to the difference of them to the $\beta $'s, too. Therefore,
the last source for the mass term could be breaking of Weyl invariance at the
classical level only. But as long as $\Phi =0$ this is excluded since both the 
action
as well as the explicit $s$-dependent Dirichlet condition in (\ref{9}) is
independent of the Weyl degree of freedom in the string world sheet metric
$\gamma $.

At this stage we can summarize: The D-brane model defined via 
(\ref{8})-(\ref{10}) is by construction equivalent to 
the open string model with free ends at the classical level. Taking into 
account string world sheet radiative corrections
this equivalence breaks down at the 1-loop order already. We do not know a
modification of the mapping of the fields and boundary conditions
which would restore T-duality.

Since a mass term is needed to model the picture described in the introduction
one should look for alternative candidates for $\sigma $-models of  
strings coupled to matrix D-branes.\footnote{The construction of \cite{mav}
seems to be of no help for our
problem since it assumes the validity of T-duality to offer the non-Abelian 
free end model as the correct description.}
Analyzing the text book dynamics of a single open string spanned in flat space
between two separated D-branes one realizes that similar to the case of
closed strings winding around a compact dimension a difference between the 
left and right moving string momentum $p_L\neq p_R $ is responsible for
the mass generation. Hence one could speculate whether the wanted 
$\sigma $-model can be found within a formalism using a split in left
and right moving contributions to the string position field along the lines of 
ref. \cite{ts}. 
\\[3mm]
\noindent
{\bf Acknowledgement} I would like to thank H.-J. Otto, V. Pershin and 
C. Preitschopf for useful discussions.

\end{document}